\documentclass{article}

\usepackage{arxiv}

\usepackage[utf8]{inputenc} % allow utf-8 input
\usepackage[T1]{fontenc}    % use 8-bit T1 fonts
\usepackage{hyperref}       % hyperlinks
\usepackage{url}            % simple URL typesetting
\usepackage{booktabs}       % professional-quality tables
\usepackage{amsfonts}       % blackboard math symbols
\usepackage{nicefrac}       % compact symbols for 1/2, etc.
\usepackage{microtype}      % microtypography
\usepackage{lipsum}
\usepackage{color}
\usepackage{algorithm}
\usepackage{algorithmic}
\usepackage{amsmath,amsfonts,amssymb}
\usepackage{subfigure}
\usepackage{lineno}
\usepackage{graphicx}
\usepackage{bm}
\usepackage{subfigure}
\usepackage{setspace}
\usepackage{graphicx}
\usepackage{newtxtext}
\usepackage{newtxmath}
\usepackage{natbib,bm}
\usepackage{siunitx}
\usepackage{amsmath,amssymb,mathtools}
\usepackage{hyperref,lipsum}
\usepackage{booktabs}
\hypersetup{
    colorlinks = true,
    urlcolor   = blue,
    citecolor  = black,
}

\newcommand{\RomanNumeralCaps}[1]
\linenumbers
\graphicspath{{./figs/}}

\usepackage{color}
\definecolor{blue}{rgb}{0, 0.4470, 0.7410}
\definecolor{red}{rgb}{0.8500, 0.1250, 0.0480} 
\definecolor{green}{rgb}{0.4660, 0.6740, 0.1880}

\title{Lift-element-based dynamic network \\ of vortex gust-airfoil interactions}

\author{
So Yamamoto and Kai Fukami$^{*}$
\\
\\
Department of Aerospace Engineering, Graduate School of Engineering, Tohoku University,\\
Sendai, Miyagi 980-8579, Japan\\
$^{*}$Corresponding author: kfukami1@tohoku.ac.jp
}

\begin{document}
\maketitle

\begin{abstract}
We construct a lift-element-based (LE) network for analyzing extreme vortex gust-airfoil interactions. The edge weights quantify the magnitude of changes in local lift elements induced by vorticity perturbations. Network broadcast mode analysis then extracts forcing and response modes that characterize input-output pathways over the evolving network.
We consider three-dimensional flows around a NACA 0012 airfoil with a Taylor vortex gust at a Reynolds number of 1000 and compare the LE modes with those of a Biot--Savart (BS) network.
The leading BS modes highlight the vortical structures formed through the gust-airfoil interactions, whereas the leading LE modes connect the gust impingement on the airfoil to the shear layer thickened by the gust, which lies on the pressure or suction side depending on the gust condition.
Perturbations along the leading BS forcing modes amplify or attenuate the gust-induced lift variation, whereas those along the leading LE forcing modes yield a sustained lift increase or decrease.
These input-output mode pairs, representing cause-and-effect relationships within the dynamic network, can guide the selection of actuation locations and timing for manipulating transient aerodynamic loads.
\end{abstract}

%%%%%%%%%%%%%%%%%%%%%%%%%%%%%%%%%%%%%%%%%%%%%%%%%%%%%%%%%%%%%%
%%%%%%%%%%%%%%%%%%%%%%%%%%%%%%%%%%%%%%%%%%%%%%%%%%%%%%%%%%%%%%
%%%%%%%%%%%%%%%%%%%%%%%%%%%%%%%%%%%%%%%%%%%%%%%%%%%%%%%%%%%%%%
\section{Introduction}

Small-scale air vehicles may encounter gust velocities comparable to their flight speeds, particularly in urban and mountainous environments~\cite{Jones2022,Fukami2023,taira2026extreme}.
A vortex gust induces local upwash or downwash as it approaches an airfoil and alters the development of its shear layers~\cite{Zhong2025}.
Strong vortex gusts can reorganize the separated flow around an airfoil and induce large, rapid variations in aerodynamic loads~\cite{Fukami2023,Fukami2024,Zhong2025,LopezDoriga2026Geometry,jones2021overview,Biler2019,Biler2021,AndreuAngulo2020,Barnes2018CCW,Barnes2018CW}.
Identifying the vortical interactions responsible for these variations is important for understanding and manipulating transient aerodynamic loads~\cite{Chen2024,Chen2026}.

Modal analysis has been widely used to extract the dominant structures and dynamic characteristics of fluid flows~\cite{Taira2017,Taira2020,fukami2026practical}.
Representative methods include data-driven approaches such as proper orthogonal decomposition (POD)~\cite{Lumley1967,Sirovich1987,Berkooz1993} and dynamic mode decomposition (DMD)~\cite{Schmid2010}, as well as operator-based approaches such as stability analysis~\cite{Noack1994} and resolvent analysis~\cite{McKeon2010,Taira2017}.
For transient flows, the base flow itself varies in time~\cite{linot2025extracting}.
Data-driven time-dependent bases have been used to track evolving vortical structures in turbulent airfoil wake-extreme gust interactions~\cite{zamani2026data}.
Space-time and wavelet-based formulations extend resolvent analysis to flows with time-varying mean states~\cite{LopezDoriga2024,Ballouz2024}.

While these methods characterize flow structures and perturbation dynamics, the extracted modes need not directly reflect aerodynamic loads.
Several studies have aimed to identify the flow structures responsible for aerodynamic forces~\cite{LiWu2018,Li2020VFM,Prakhar2025Vortices}.
Modal force partitioning quantifies the aerodynamic load contributions of decomposed flow modes~\cite{Prakhar2025}.
Information-theoretic machine learning has been used to extract time-varying informative vortical structures that contribute to the future evolution of aerodynamic coefficients~\cite{FukamiAraki2026,Koshikawa2026Causal,Koshikawa2026Gust}.

Network analysis offers an interaction-based representation of fluid flows, in which edge weights can be defined according to the physics~\cite{Newman2018, Taira2022}.
In fluid mechanics, network-based approaches have been developed using vortical interactions~\cite{Nair2015,Taira2016}, modal interactions~\cite{Nair2018}, Lagrangian particle trajectories~\cite{Ser2015}, and time-series measurements~\cite{Iacobello2018}.
To characterize interactions in time-varying networks, dynamic communicability extends the Katz centrality~\cite{Katz1953} by accumulating time-ordered walks over evolving networks~\cite{Grindrod2011,grindrod2013matrix,Grindrod2014}.
Singular value decomposition of the communicability matrix yields forcing and response modes that describe network inputs and outputs.
Since the network is constructed from the interactions in the flow, these modes identify where perturbations can be introduced and where their effects are received, providing guidance for flow control~\cite{Yeh2021,Patel2026}.

Physical interactions captured by a network depend on the definition of its edge weights.
Biot--Savart networks quantify vortical interactions through induced velocities~\cite{Nair2015,Taira2022} and have revealed the scale-free structure of two-dimensional isotropic turbulence~\cite{Taira2016}.
Navier--Stokes networks instead quantify how a vorticity pulse at one node changes the right-hand side of the vorticity transport equation at another~\cite{Yeh2021}.
Other edge weights have been defined by the energy transfer between modes~\cite{Nair2018}, the proximity of Lagrangian particle trajectories~\cite{Hadjighasem2016,Iacobello2019}, and, for sensor measurements, the visibility, recurrence, and cluster transitions of time series~\cite{Murugesan2015,Godavarthi2017,Nair2019}.
However, none of these edge weights directly quantify changes in aerodynamic loads.
A network formulation that incorporates these load changes is therefore of interest.

To incorporate load-related interactions, we introduce a force-element-based network whose edge weights quantify the magnitude of changes in local force elements induced by vorticity perturbations.
Force-element analysis decomposes aerodynamic forces into volume and surface contributions~\cite{Quartapelle1983,Chang1992}.
Focusing on the lift element, we construct time-evolving BS and lift-element-based (LE) networks for three-dimensional extreme vortex gust-airfoil interactions around a NACA 0012 airfoil at a Reynolds number of 1000, and compare their forcing and response modes to identify the regions emphasized by each network and their relation to the transient lift.
We further introduce perturbations along the leading forcing modes into the flow to examine their effects on the transient lift.

The definitions of the two networks and the network broadcast mode analysis are introduced in Sec.~\ref{sec:method}.
The flow physics of the vortex gust encounters considered in this study is described in Sec.~\ref{sec:flow}.
The modal structures and the perturbation analysis are discussed in Sec.~\ref{sec:results}.
Concluding remarks are in Sec.~\ref{sec:conclusion}.

%%%%%%%%%%%%%%%%%%%%%%%%%%%%%%%%%%%%%%%%%%%%%%%%%%%%%%%%%%%%%%
%%%%%%%%%%%%%%%%%%%%%%%%%%%%%%%%%%%%%%%%%%%%%%%%%%%%%%%%%%%%%%
%%%%%%%%%%%%%%%%%%%%%%%%%%%%%%%%%%%%%%%%%%%%%%%%%%%%%%%%%%%%%%
\section{Networked modal analysis for aerodynamic flows}\label{sec:method}
\begin{figure}[!t]
    \centering
    \includegraphics[width=164.9mm]{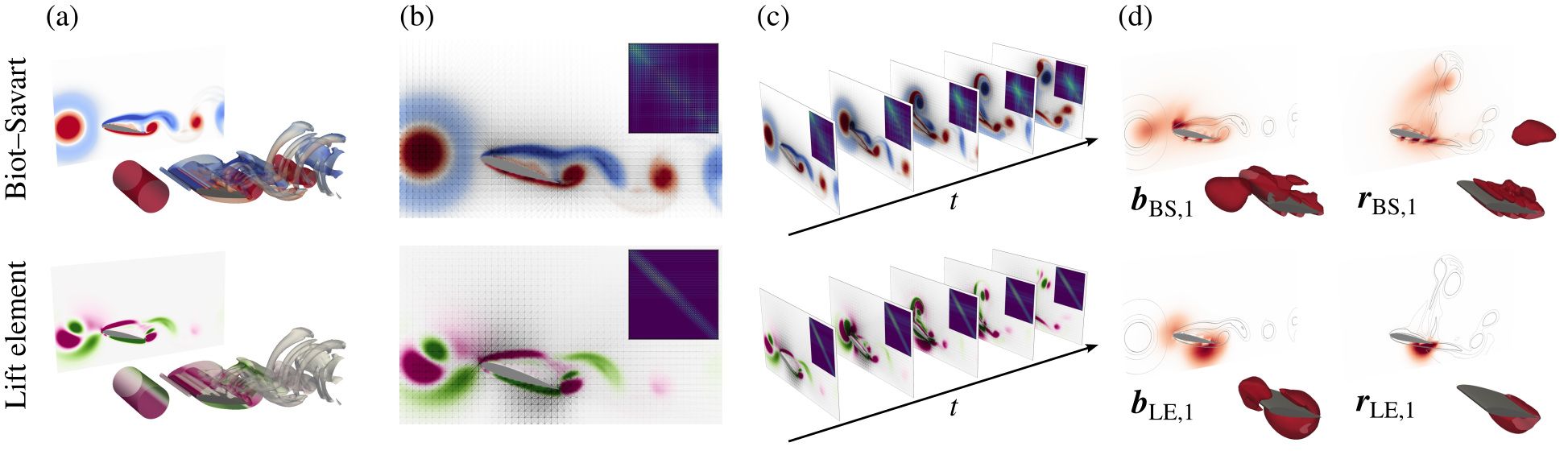}
    \caption{Overview of the present network broadcast mode analysis for the Biot--Savart (BS, top row) and lift-element-based (LE, bottom row) networks.
    (a) Three-dimensional flow field, $Q$-criterion isosurfaces colored by the spanwise vorticity $\omega_z$ (top) and by the lift element $L_E$ (bottom).
    (b) Network over the flow field with the corresponding adjacency matrix $\bm{A}_k$.
    (c) Time-evolving network over the gust encounter for $(G, D/c) = (2, 0.5)$.
    (d) Leading forcing modes $\bm{b}_{\mathrm{BS},1}$, $\bm{b}_{\mathrm{LE},1}$ and response modes $\bm{r}_{\mathrm{BS},1}$, $\bm{r}_{\mathrm{LE},1}$.}
    \label{fig:overview}
\end{figure}

To identify how local vortical structures affect the transient lift in extreme vortex gust-airfoil interactions, we construct a lift-element-based network and analyze it through network broadcast mode analysis, as summarized in Fig.~\ref{fig:overview}.
The flow field at each instant is represented as a weighted network, as shown in Fig.~\ref{fig:overview}(a,b), which evolves over the gust encounter, as visualized in Fig.~\ref{fig:overview}(c).
The forcing and response modes are obtained from the singular value decomposition of the communicability matrix of this time-evolving network, as depicted in Fig.~\ref{fig:overview}(d).
We first define the BS and LE edge weights in Sec.~\ref{sec:definitions} and then describe the time-evolving network analysis in Sec.~\ref{sec:tenet}.
\subsection{Network definitions}\label{sec:definitions}
\subsubsection{Biot--Savart network}

Let us first consider the Biot--Savart (BS) network, whose edge weights are based on the velocities induced between vortical elements.
For grid cells of volume $\Delta V$, the magnitude of the velocity induced at $\bm{x}_i$ by the vortical element at $\bm{x}_j$ is given by~\cite{Gopalakrishnan2021}
\begin{equation}
  u_{ij} = \frac{\left|\bm{\omega}(\bm{x}_j, t) \times (\bm{x}_i - \bm{x}_j)\right| \Delta V}{4\pi \left|\bm{x}_i - \bm{x}_j\right|^{3}} \quad (i \neq j).
  \label{eq:bs_u}
\end{equation}
The edge weight is defined as the average of the velocities induced by the two elements on each other~\cite{Nair2015,Taira2016},
\begin{equation}
  A_{\mathrm{BS},ij} = \frac{u_{ij} + u_{ji}}{2},
  \label{eq:bs}
\end{equation}
with $A_{\mathrm{BS},ii} = 0$ since a vortical element induces no velocity on itself.
The adjacency matrix $\bm{A}_{\mathrm{BS}}$ is therefore symmetric and represents the mutual induction between vortical elements.
This network has been mostly used for network-based analyses in fluid mechanics and combustion~\cite{Taira2022,murayama2018characterization}.
Since the edge weight increases with the vorticity of the two elements, the BS network emphasizes vortical structures with strong vorticity, rather than their contribution to the lift.

%%%%%%%%%%%%%%%%%%%%%%%%%%%%%%%%%%%%%%%%%%%%%%%%%%%%%%%%%%%%%%
%%%%%%%%%%%%%%%%%%%%%%%%%%%%%%%%%%%%%%%%%%%%%%%%%%%%%%%%%%%%%%
\subsubsection{Lift-element-based network}

To identify the vortical regions that affect the lift, which the BS network may not directly extract, we propose the second network based on the lift element.
We introduce an auxiliary potential $\phi$ satisfying $\nabla^2\phi = 0$ with $\partial\phi/\partial n = -n_y$ on the airfoil surface, where $\bm{n}$ is the unit normal directed from the airfoil into the fluid.
For the unit nondimensional span and vanishing far-field boundary contributions, the nondimensional lift can be decomposed as~\cite{Quartapelle1983,Chang1992,Koshikawa2026Gust}
\begin{equation}
    F_L = \int_V(\bm{\omega}\times\bm{u})\cdot\nabla\phi\,\mathrm{d}V  + \frac{1}{Re}\int_S (\bm{\omega}\times\bm{n})\cdot(\nabla\phi+\bm{e}_y)\,\mathrm{d}S,
  \label{eq:lift_decomposition}
\end{equation}
where $V$ is the exterior fluid volume over one spanwise period, $S$ is the airfoil surface over the same period, and $\bm{e}_y$ is the unit vector in the lift direction.
The first and second terms represent the volume and surface contributions, respectively.
The integrand of the volume term, referred to as the lift element, is
\begin{equation}
  L_E(\bm{\omega}, \bm{u}) = \left(\bm{\omega} \times \bm{u}\right) \cdot \nabla\phi,
  \label{eq:le}
\end{equation}
which represents the local contribution of the interaction between vorticity and velocity to the lift.

The second edge weight in this study considers the change of the lift element at $\bm{x}_i$ due to a localized pulse introduced at $\bm{x}_j$,
\begin{equation}
  A_{\mathrm{LE},ij} = \frac{\big| L_E(\bm{\omega} + \epsilon\bm{\delta}_j, \bm{u} + \bm{u}'_j) - L_E(\bm{\omega}, \bm{u}) \big|_{\bm{x}_i}}{\epsilon},
  \label{eq:le_net}
\end{equation}
where $\bm{u}'_j$ is the velocity induced by the vorticity pulse at $\bm{x}_j$,
\begin{equation}
  \epsilon\bm{\delta}_j = \frac{\epsilon}{2 r_\delta^2} \Big[ \left( 2 r_\delta^2 - r_j^2 \right) \bm{e}_\delta + (\bm{r}_j \cdot \bm{e}_\delta)\, \bm{r}_j \Big] \exp\left( -\frac{r_j^2}{2 r_\delta^2} \right),
  \label{eq:pulse}
\end{equation}
where $\bm{r}_j = \bm{x} - \bm{x}_j$, $r_j = |\bm{r}_j|$ and $\bm{e}_\delta = (1, 1, 1)/\sqrt{3}$ is the axis of the pulse.
The pulse has a vorticity of $\epsilon\bm{e}_\delta$ at its center, so that all three vorticity components are perturbed.
We use $\epsilon c/U_\infty = 0.001$ and $r_\delta = 0.12c$, corresponding to 1.5 times the network grid spacing, so that the pulse is sufficiently weak and localized while its influence on the lift element is resolved.
The pulse decays rapidly away from $\bm{x}_j$, but its effect on the lift elements depends on the local vorticity, velocity, and auxiliary-potential gradient.
Indirect influences on more distant regions are captured through walks over the time-evolving network.
Because of the absolute value in Eq.~(\ref{eq:le_net}), the LE network retains the magnitudes of lift-element changes, and the modal signs do not indicate the direction of lift change.
The $n$-th modes of the BS and LE networks are denoted by $(\bm{b}_{\mathrm{BS},n}, \bm{r}_{\mathrm{BS},n})$ and $(\bm{b}_{\mathrm{LE},n}, \bm{r}_{\mathrm{LE},n})$.

%%%%%%%%%%%%%%%%%%%%%%%%%%%%%%%%%%%%%%%%%%%%%%%%%%%%%%%%%%%%%%
%%%%%%%%%%%%%%%%%%%%%%%%%%%%%%%%%%%%%%%%%%%%%%%%%%%%%%%%%%%%%%
\begin{figure}[!t]
    \centering
    \includegraphics[width=164.5mm]{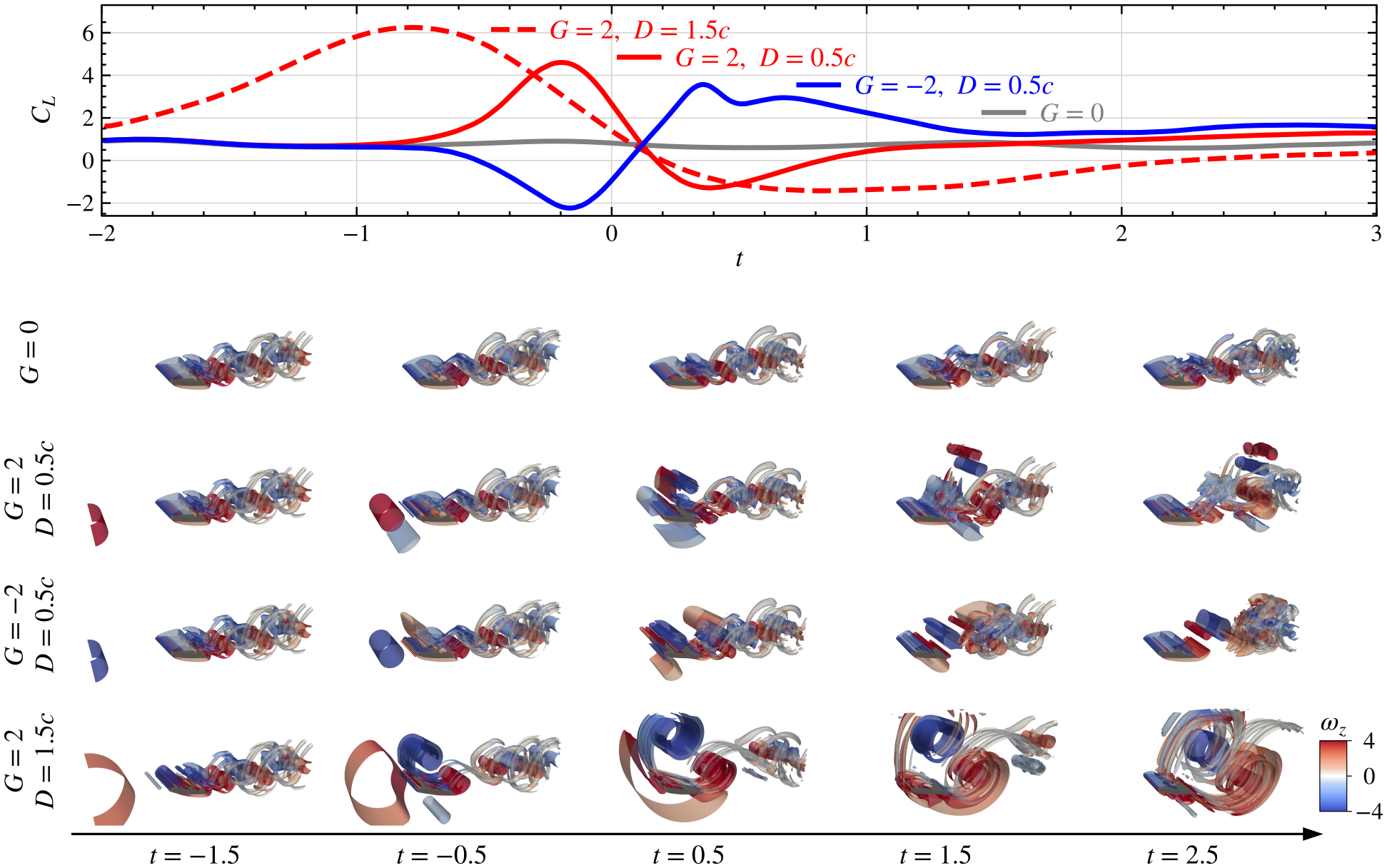}
    \caption{Extreme vortex gust-airfoil interaction for the three gust conditions, together with the undisturbed flow ($G = 0$). Top: lift coefficient $C_L$ for $G = 0$ and $(G, D/c) = (2, 0.5)$, $(-2, 0.5)$, and $(2, 1.5)$. Bottom: $Q$-criterion isosurfaces colored by the spanwise vorticity $\omega_z$ for $-1.5 \le t \le 2.5$ for the four cases.}
    \label{fig:cl_three}
\end{figure}

\subsection{Time-evolving network}\label{sec:tenet}

Each grid cell at $\bm{x}_i$ is treated as a network node.
The resulting network of $N$ nodes is described by the adjacency matrix $\bm{A}\in\mathbb{R}^{N\times N}$, whose entry $A_{ij}$ quantifies the influence of node $j$ on node $i$.
The network response to a perturbation $\bm{f}\in\mathbb{R}^N$ distributed over the nodes is modeled as
\begin{equation}
  \bm{q} = \left(\bm{I} + \alpha\bm{A} + \alpha^2\bm{A}^2 + \cdots\right)\bm{f} = \bm{K}\bm{f},
  \label{eq:katz_series}
\end{equation}
where $\bm{q}\in\mathbb{R}^N$ is the network response and $\alpha$ is the walk-downweighting parameter.
For $\alpha < 1/\rho(\bm{A})$, where $\rho$ is the spectral radius, the series converges to the Katz function~\cite{Katz1953}
\begin{equation}
  \bm{K} = \left(\bm{I} - \alpha\bm{A}\right)^{-1}.
  \label{eq:katz}
\end{equation}

For a time-evolving network, the adjacency matrix is sampled at discrete times $t_k$ $(k = 0, 1, \dots, m)$ as $\bm{A}_k$, which is considered time-frozen until the next sample.
A perturbation can then propagate from node $j$ to node $i$ through a time-respecting walk~\cite{Grindrod2011}.
The spread over $t \in [t_0, t_m]$ is thus given by the product of the Katz functions in temporal order,
\begin{equation}
  \bm{q}(t_m) = \bm{K}_m \bm{K}_{m-1} \cdots \bm{K}_1 \bm{K}_0\, \bm{f}(t_0) = \bm{S}_m \bm{f}(t_0),
  \label{eq:chain}
\end{equation}
where $\bm{K}_k = (\bm{I} - \alpha\bm{A}_k)^{-1}$ is the Katz function of $\bm{A}_k$, and $\bm{S}_m$ is the dynamic communicability matrix~\cite{Grindrod2011}.
The entry $S_{m,ij}$ quantifies the influence of a perturbation introduced at node $j$ at $t_0$ on node $i$ at $t_m$ through all time-respecting walks.

To account for the decay of information over time, age downweighting generalizes Eq.~(\ref{eq:chain}) to the recurrence~\cite{grindrod2013matrix}
\begin{equation}
  \bm{S}_{k+1} = \bm{K}_{k+1} \left[\bm{I} + e^{-\gamma\Delta t}\left(\bm{S}_k - \bm{I}\right)\right]
  \label{eq:comm}
\end{equation}
for $k = 0, 1, \dots, m-1$ with $\bm{S}_0 = \bm{K}_0$, where $\gamma$ is the age-downweighting parameter and $\Delta t$ is the sampling interval.
Equation~(\ref{eq:comm}) recovers Eq.~(\ref{eq:chain}) for $\gamma = 0$.
In the limit $\gamma\to\infty$, $\bm{S}_m\to\bm{K}_m$, so the modes depend only on the final snapshot.

All quantities are nondimensionalized by $c$ and $U_\infty$.
The sampling interval $\Delta t = 0.01$ corresponds to the interval between flow snapshots.
The walk-downweighting parameter is set to $\alpha = \Delta t$, for which the Katz function corresponds to a backward Euler step of the network dynamics $\mathrm{d}\bm{q}/\mathrm{d}t = \bm{A}\bm{q}$ over $\Delta t$.
The age-downweighting parameter $\gamma = U_\infty/c$, i.e., unity in the nondimensional units, combines the network evolution with the decay of earlier interactions over the convective time scale.

The forcing and response modes are obtained from the singular value decomposition of the communicability matrix~\cite{Yeh2021},
\begin{equation}
  \bm{S}_m = \bm{R}\,\bm{\Sigma}\,\bm{B}^{\top} = \sum_{n=1}^{N} \sigma_n\, \bm{r}_n \bm{b}_n^{\top},
  \label{eq:svd}
\end{equation}
where the forcing modes $\bm{b}_n$ and the response modes $\bm{r}_n$ are the orthonormal right- and left-singular vectors, respectively, and the singular values satisfy $\sigma_1 \ge \sigma_2 \ge \cdots \ge 0$ and $\bm{S}_m\bm{b}_n = \sigma_n\bm{r}_n$.
The leading forcing mode maximizes the network response norm for a unit-norm input, with gain $\sigma_1$.
The forcing mode identifies regions where perturbations effectively spread over the time-evolving network, while the response mode identifies regions that receive their effects.
The overall sign of each forcing-response mode pair is arbitrary.

%%%%%%%%%%%%%%%%%%%%%%%%%%%%%%%%%%%%%%%%%%%%%%%%%%%%%%%%%%%%%%
%%%%%%%%%%%%%%%%%%%%%%%%%%%%%%%%%%%%%%%%%%%%%%%%%%%%%%%%%%%%%%
%%%%%%%%%%%%%%%%%%%%%%%%%%%%%%%%%%%%%%%%%%%%%%%%%%%%%%%%%%%%%%
\section{Transient flow physics of vortex gust-airfoil interaction}\label{sec:flow}

Here, let us examine the flow of interest in this study.
We consider three-dimensional extreme vortex gust-airfoil interactions around a NACA 0012 airfoil at a chord-based Reynolds number of $Re = U_\infty c/\nu = 1000$ and an angle of attack of $18^\circ$.
At this condition, the flow is dominated by large two-dimensional vortical structures, while exhibiting three-dimensional structures~\cite{kouser2021direct, Gupta_Zhao_Sharma_Agrawal_Hourigan_Thompson_2023}.

Direct numerical simulations of the incompressible Navier--Stokes equations are performed using the spectral-element solver Nek5000~\cite{Nek5000}.
The computational domain extends over $(x, y, z)/c \in [-20,23] \times [-20,20] \times [0,1]$ with the leading edge at the origin.
A uniform velocity $U_\infty$ is prescribed at the inlet, a symmetry condition at $y/c = \pm 20$, the standard outflow condition of Nek5000 at the outlet, a no-slip condition on the airfoil surface, and a periodic condition in the spanwise direction.
Time integration uses the second-order backward differentiation scheme with third-order extrapolation of the advection term.
Verification and validation have been performed for both the undisturbed flow and the vortex gust-airfoil interactions in the same manner as in previous studies~\cite{Fukami2023,Fukami2024,LopezDoriga2026Geometry}.
The lift coefficient is defined as $C_L = F_L/(0.5\rho U_\infty^2 c)$, where $F_L$ is the lift force per unit span and $\rho$ is the density.

\begin{figure}[!t]
    \centering
    \includegraphics[width=165.0mm]{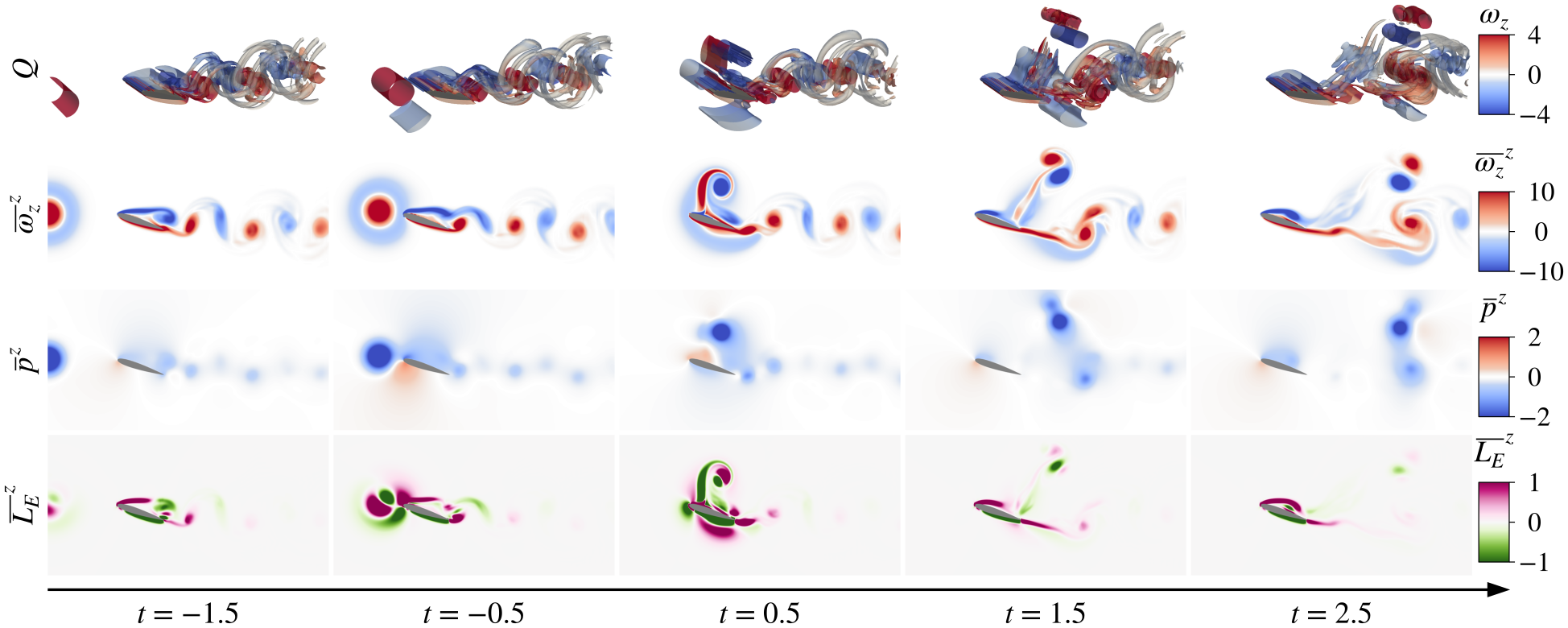}
    \caption{Flow field for $(G, D/c) = (2, 0.5)$ for $-1.5 \le t \le 2.5$. From top to bottom: three-dimensional vortical structures ($Q$-criterion isosurfaces colored by $\omega_z$), and spanwise-averaged spanwise vorticity $\overline{\omega_z}^{\,z}$, pressure $\overline{p}^{\,z}$, and lift element $\overline{L_E}^{\,z}$.}
    \label{fig:g2_d0.5_flow}
\end{figure}

A vortex gust is introduced upstream of the airfoil at $x_0/c = -2$ and $y_0/c = 0$ by superposing its velocity field on that of the baseline flow.
The disturbance is modeled as a spanwise-uniform Taylor vortex with rotational velocity profile~\cite{Fukami2023,Fukami2024,fukami2025extreme,Koshikawa2026Gust}
\begin{equation}
  u_\theta = u_{\theta,\max}\frac{r}{R}\exp\left[\frac{1}{2}\left(1-\frac{r^2}{R^2}\right)\right],
  \label{eq:taylor_gust}
\end{equation}
where $r$ is the radial distance from the vortex center and $R$ is the radius of maximum azimuthal speed $|u_{\theta,\max}|$.
The vortex gust is characterized by the gust ratio $G \equiv u_{\theta,\max}/U_\infty$, with positive values for counter-clockwise rotation, and its size $D \equiv 2R$.

We examine $(G, D/c) = (2, 0.5)$, $(-2, 0.5)$, and $(2, 1.5)$.
Here, $|G| > 1$ corresponds to extreme aerodynamic conditions where the gust velocity exceeds the free-stream velocity~\cite{taira2026extreme,Fukami2023,jones2021overview}.
Comparing the positive and negative gusts of the same size reveals how the gust rotation changes the transient lift, and comparing the two positive gusts reveals the effect of gust size.
The convective time $t$ is set to zero when the center of the vortex arrives at the leading edge of the airfoil.

\begin{figure}[!t]
    \centering
    \includegraphics[width=165.0mm]{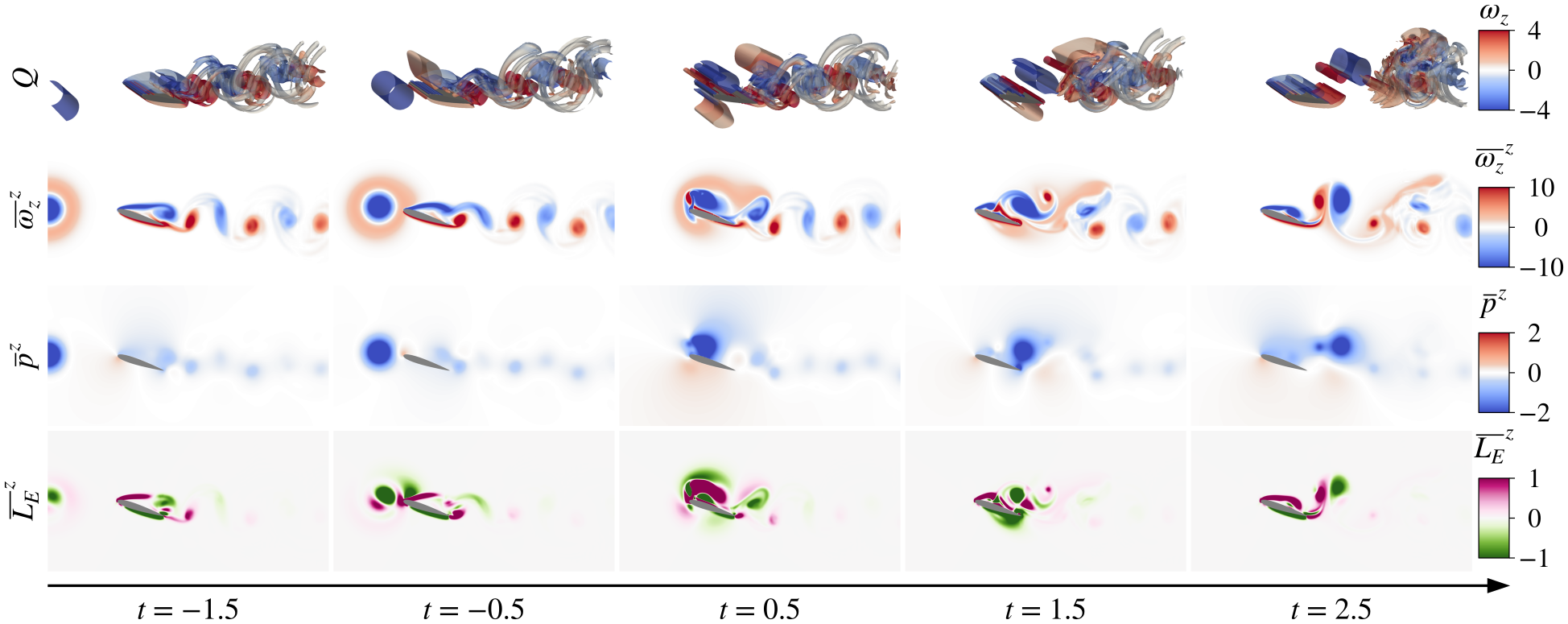}
    \caption{Flow field for $(G, D/c) = (-2, 0.5)$ for $-1.5 \le t \le 2.5$. From top to bottom: three-dimensional vortical structures ($Q$-criterion isosurfaces colored by $\omega_z$), and spanwise-averaged spanwise vorticity $\overline{\omega_z}^{\,z}$, pressure $\overline{p}^{\,z}$, and lift element $\overline{L_E}^{\,z}$.}
    \label{fig:g-2_d0.5_flow}
\end{figure}

\begin{figure}[!t]
    \centering
    \includegraphics[width=165.0mm]{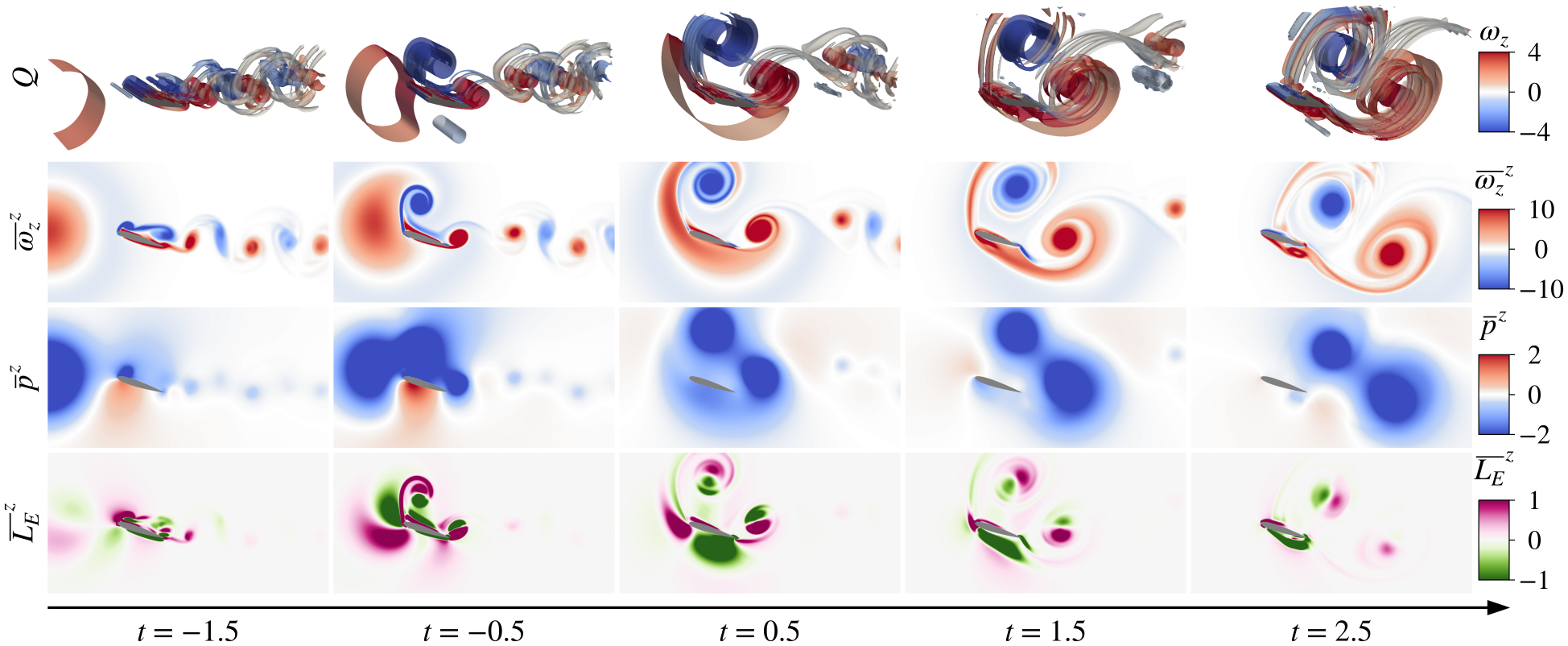}
    \caption{Flow field for $(G, D/c) = (2, 1.5)$ for $-1.5 \le t \le 2.5$. From top to bottom: three-dimensional vortical structures ($Q$-criterion isosurfaces colored by $\omega_z$), and spanwise-averaged spanwise vorticity $\overline{\omega_z}^{\,z}$, pressure $\overline{p}^{\,z}$, and lift element $\overline{L_E}^{\,z}$.}
    \label{fig:g2_d1.5_flow}
\end{figure}

The time histories of $C_L$ and the corresponding three-dimensional vortical structures for the three gust conditions are presented in Fig.~\ref{fig:cl_three}.
For the positive small gust, the lift first increases and then decreases as the pair of the gust and the LEV departs from the airfoil.
Reversing the gust rotation produces an initial lift decrease followed by an overshoot as the gust merges with the LEV over the suction side.
Increasing the size of the positive gust produces an earlier and broader lift maximum followed by a prolonged lift reduction, accompanied by the formation and departure of a large pair of the leading-edge-vortex (LEV) and the trailing-edge-vortex (TEV).
Figures~\ref{fig:g2_d0.5_flow}--\ref{fig:g2_d1.5_flow} show the three-dimensional vortical structures for each case together with the spanwise-averaged spanwise vorticity $\overline{\omega_z}^{\,z}$, pressure $\overline{p}^{\,z}$, and lift element $\overline{L_E}^{\,z}$.
Positive $\omega_z$ corresponds to counter-clockwise rotation.

The undisturbed baseline flow exhibits alternating shedding of clockwise LEVs from the suction side and counter-clockwise TEVs.
The cores of the LEV and TEV are nearly uniform along the span, whereas the wake contains three-dimensional secondary structures.

For $(G, D/c) = (2, 0.5)$, as seen in Fig.~\ref{fig:g2_d0.5_flow}, the counter-clockwise gust approaching the leading edge ($t = -0.5$) induces an upwash.
A region of high pressure forms below the leading edge, while the low-pressure core of the gust lies ahead of the suction side.
Positive lift elements surround the gust and the leading edge.
The LEV grows, and the lift reaches its maximum.
The gust and the LEV then form a counter-rotating pair that lifts off from the suction side ($t = 0.5$).
The low-pressure region associated with the vortex pair moves away from the airfoil surface.
Negative lift elements occupy the region around the pair near the leading edge, and the lift reaches its minimum.
Meanwhile, the pressure-side shear layer thickens and rolls up into a TEV near the trailing edge ($t = 1.5$), where positive lift elements appear along the pressure side and behind the trailing edge as the lift recovers.
At $t = 2.5$, the pair has moved away from the airfoil and a new LEV forms at the leading edge, recovering a distribution of lift elements similar to that at $t = -1.5$.

For $(G, D/c) = (-2, 0.5)$, as seen in Fig.~\ref{fig:g-2_d0.5_flow}, the clockwise gust induces a downwash, and the lift decreases.
The gust then merges with the LEV, which rotates in the same direction, into a large LEV with a thick shear layer on the suction side ($t = 0.5$).
The low-pressure core of the merged LEV lies over the forward suction surface, and an arc of positive lift elements around the merged LEV coincides with the lift overshoot, while negative lift elements appear inside the LEV core and below the leading edge.
The merged LEV is advected along the suction side ($t = 1.5$), with its low-pressure core moving toward the trailing edge, and is shed together with a TEV at $t \approx 2.5$.
During the interaction, the pressure-side shear layer remains thin and the TEV is weak.
The merged LEV remains a coherent column along the span.

For $(G, D/c) = (2, 1.5)$, as seen in Fig.~\ref{fig:g2_d1.5_flow}, the larger gust carries a broad low-pressure region and induces a prolonged upwash.
A large region of high pressure forms below the leading edge, and a large LEV forms on the suction side, with the lift reaching its maximum.
The LEV leaves the airfoil at $t \approx -0.5$, followed by a large TEV that rolls up at the trailing edge at $t \approx 0.5$; the two vortices appear as separate low-pressure cores above and behind the airfoil.
The pressure-side shear layer thickens upstream to approximately mid-chord, the pressure below the airfoil decreases, and large negative lift elements spread below the airfoil, coinciding with the lift minimum.
The two large vortices remain coherent along the span and leave the airfoil as a pair ($t = 1.5$ and $2.5$), and no periodic shedding is observed during the interaction.

For network analysis, we set the analysis window to $-0.7 \le t \le 1.2$ for $D = 0.5c$ and $-2 \le t \le 2$ for $D = 1.5c$ to cover the lift maximum and minimum during the gust encounter, with a longer window for the larger gust, which reaches the airfoil earlier and prolongs the lift variation.
The networks are built on a uniform grid with spacing $0.08c$ over $(x, y, z)/c \in [-1.5,4.5] \times [-0.75,1.5] \times [0,1]$.
This domain includes the region upstream of the leading edge where the gust approaches, the airfoil, and the near wake into which the gust, LEVs, and TEVs are advected.
Here, it captures the vortical interactions that affect the lift during the analysis windows.
%%%%%%%%%%%%%%%%%%%%%%%%%%%%%%%%%%%%%%%%%%%%%%%%%%%%%%%%%%%%%%
%%%%%%%%%%%%%%%%%%%%%%%%%%%%%%%%%%%%%%%%%%%%%%%%%%%%%%%%%%%%%%
%%%%%%%%%%%%%%%%%%%%%%%%%%%%%%%%%%%%%%%%%%%%%%%%%%%%%%%%%%%%%%
\section{Results and discussion}\label{sec:results}
\begin{figure}[!t]
    \centering
    \includegraphics[width=164.9mm]{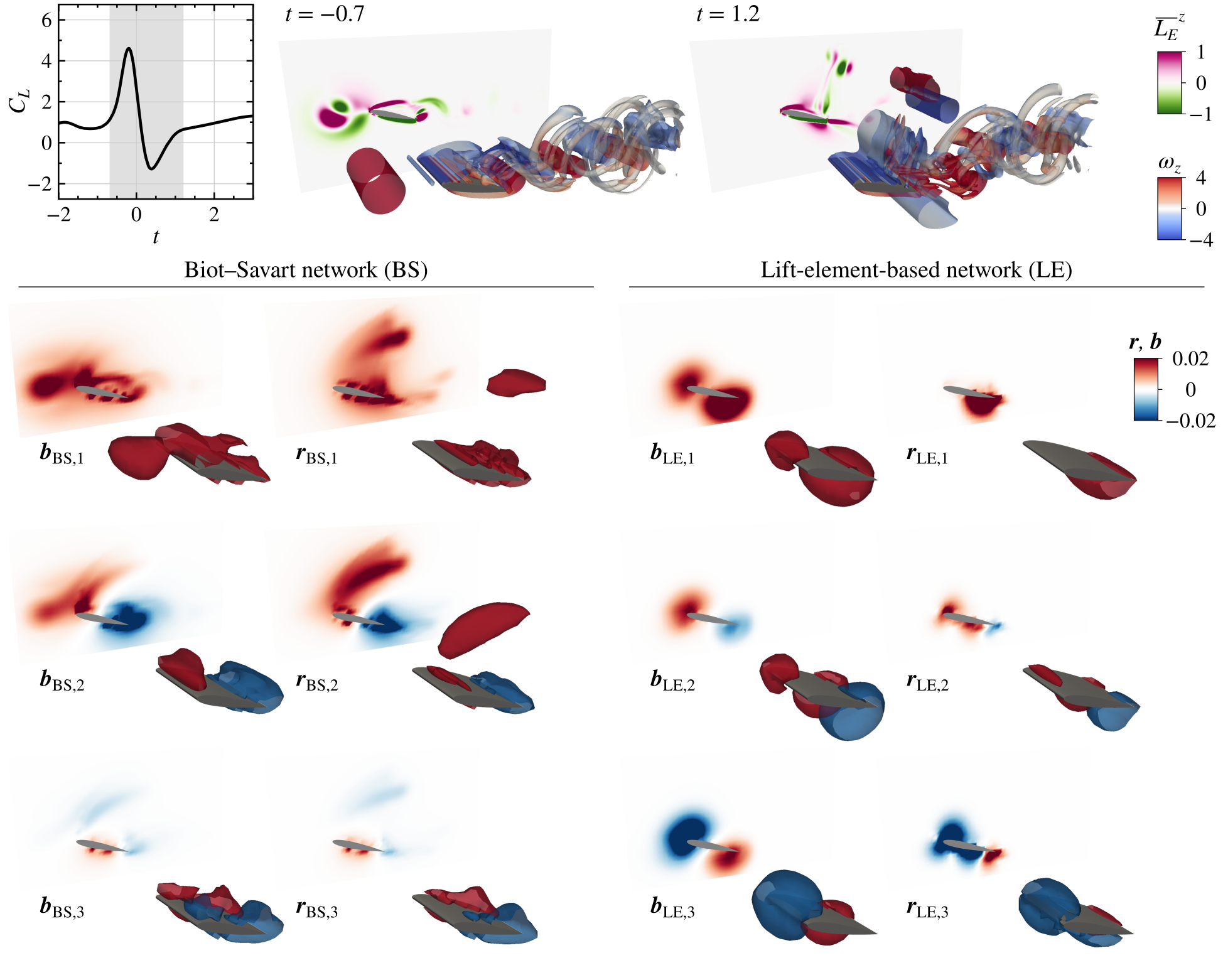}
    \caption{First three modes for $(G, D/c) = (2, 0.5)$.
    Top left: $C_L$, with the analysis window $-0.7 \le t \le 1.2$ shaded.
    Top right: flow fields at $t=-0.7$ and $1.2$ ($Q$-criterion isosurfaces colored by $\omega_z$; $\overline{L_E}^{\,z}$ on the plane).
    Bottom: forcing modes $\bm{b}_{\mathrm{BS},n}$, $\bm{b}_{\mathrm{LE},n}$ and response modes $\bm{r}_{\mathrm{BS},n}$, $\bm{r}_{\mathrm{LE},n}$ for $n=1$\textmd{--}$3$ (top to bottom), shown by spanwise averages on the plane and isosurfaces at $\pm 0.02$.}
    \label{fig:g2_d0.5_modes}
\end{figure}
Let us present the forcing and response modes of the BS and LE networks for the three vortex gust-airfoil interactions.
We first compare the modal structures of the two networks in Sec.~\ref{sec:modal} to examine which regions each network emphasizes and how the LE modes relate to the transient lift.
We then introduce perturbations along the leading forcing modes into the flow in Sec.~\ref{sec:perturbation} to assess their effects on the lift.

\subsection{Modal structure}\label{sec:modal}
Let us first examine the forcing and response modes of the BS and LE networks for the three gust conditions.
For each case, the leading modes of the two networks are discussed first, followed by the higher modes.

\subsubsection{Positive small gust: $(G, D/c) = (2, 0.5)$}
Let us present the first three mode pairs for the positive small gust in Fig.~\ref{fig:g2_d0.5_modes}.
The dominant BS modes emphasize the development and departure of the gust-LEV pair.
The leading forcing mode spreads over the incoming gust and the vortical structures around the airfoil, whereas the response mode is concentrated on the gust-LEV pair that is formed by their interactions and detaches from the airfoil.
The gust and the LEV rotate in opposite directions, and their mutual induction carries the pair away from the airfoil.

The LE network extracts a different input-output pathway.
The leading forcing mode appears at the leading edge, where the gust impinges, and below the rear pressure side, whereas the response mode is concentrated below the rear pressure side near the trailing edge.
As seen in Fig.~\ref{fig:g2_d0.5_flow}, the counter-clockwise gust strengthens the pressure-side shear layer of the same sign, which grows along the pressure side and rolls up into the TEV, and positive lift elements are distributed in this region at the end of the analysis window.
The LE pathway thus represents how the gust impingement on the leading edge generates lift through the development of the pressure-side shear layer, rather than through the departing vortex pair emphasized by the BS network.

The higher modes distinguish different vortex-formation regions.
The gust-LEV pair and the rear pressure-side region appear with opposite signs in the second BS mode, separating the input-output relationships associated with the departing vortex pair and the TEV formation.
The leading-edge region and the rear pressure-side region near the trailing edge appear with opposite signs in the third LE mode, separating the regions associated with the LEV and TEV development.
Since the networks include interactions along the span, signs can also change in the spanwise direction around the principal near-airfoil structures, as in the third BS mode and the second LE mode, which are antisymmetric about mid-span, reflecting the three-dimensionality of the flow seen in Fig.~\ref{fig:cl_three}.
Thus, the BS modes emphasize the departing gust-LEV pair, whereas the LE modes highlight the pressure-side shear layer associated with lift generation.

\begin{figure}[!t]
    \centering
    \includegraphics[width=164.9mm]{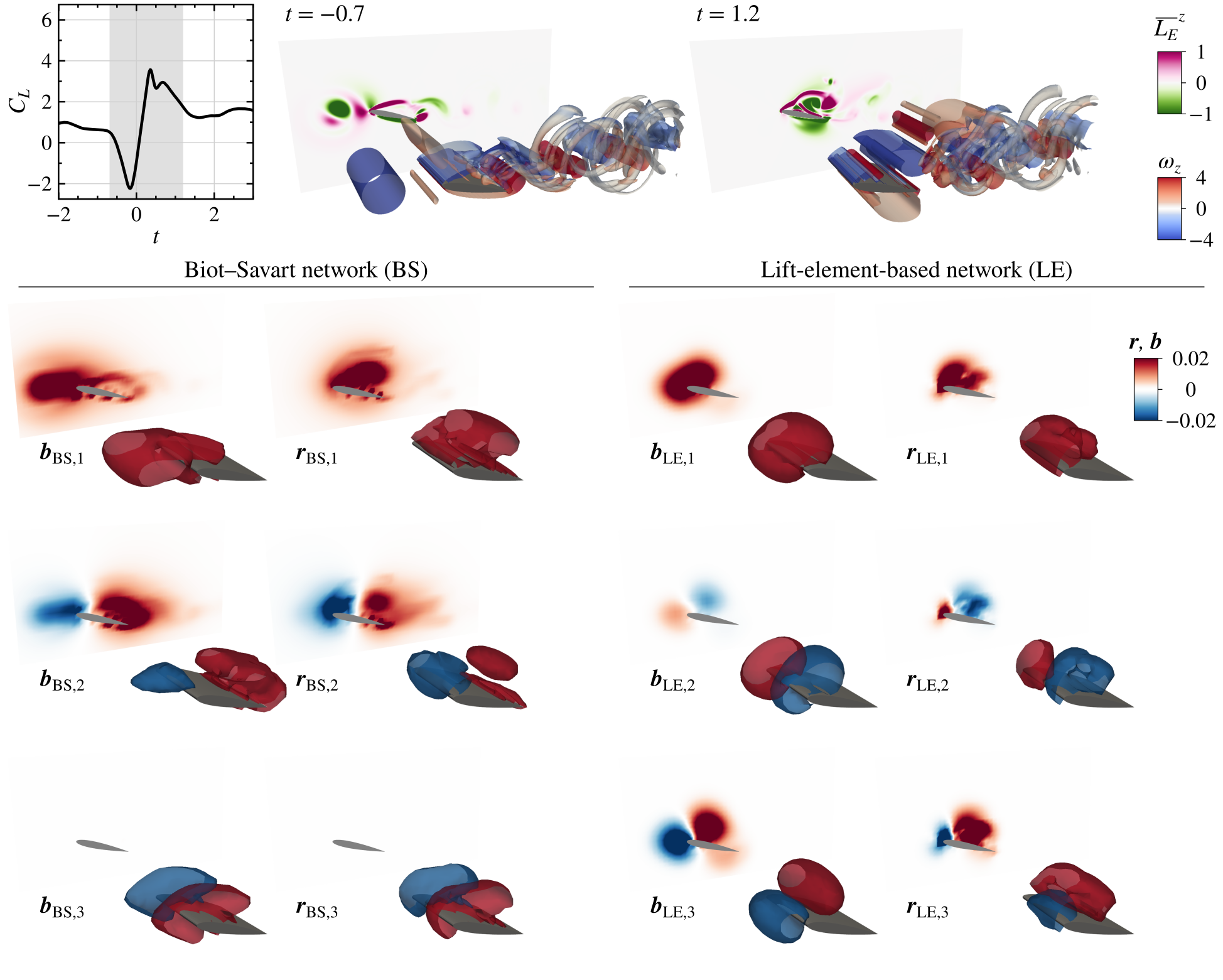}
    \caption{First three modes for $(G, D/c) = (-2, 0.5)$.
    Top left: $C_L$, with the analysis window $-0.7 \le t \le 1.2$ shaded.
    Top right: flow fields at $t=-0.7$ and $1.2$ ($Q$-criterion isosurfaces colored by $\omega_z$; $\overline{L_E}^{\,z}$ on the plane).
    Bottom: forcing modes $\bm{b}_{\mathrm{BS},n}$, $\bm{b}_{\mathrm{LE},n}$ and response modes $\bm{r}_{\mathrm{BS},n}$, $\bm{r}_{\mathrm{LE},n}$ for $n=1$\textmd{--}$3$ (top to bottom), shown by spanwise averages on the plane and isosurfaces at $\pm 0.02$.}
    \label{fig:g-2_d0.5_modes}
\end{figure}

\subsubsection{Negative small gust: $(G, D/c) = (-2, 0.5)$}

We next reverse the gust rotation while keeping its size.
Let us present the mode pairs for the negative small gust in Fig.~\ref{fig:g-2_d0.5_modes}.
For the BS network, the leading forcing mode covers the incoming gust and the leading-edge region, whereas the response mode is concentrated on the merged LEV over the suction side.
The clockwise gust and the LEV rotate in the same direction and merge into a large vortex, in contrast to the departing counter-rotating pair discussed above.

For the LE network, the leading forcing mode is concentrated around the leading edge, and the response mode lies on the suction side near the leading edge, along the shear layer surrounding the merged LEV.
The clockwise gust strengthens the suction-side shear layer of the same sign, instead of the pressure-side one.
At the end of the analysis window, positive lift elements form an arc around the merged LEV and a layer on the forward suction surface, at the location of the response mode.
The principal LE pathway thus shifts from the pressure side to the suction side with the reversal of the gust rotation, following the side on which the shear layer develops.

The higher modes distinguish the gust, the merged LEV, and the surrounding shear layer.
The upstream gust region and the merged LEV on the suction side appear with opposite signs in both the forcing and response of the second BS mode, distinguishing the gust from the suction-side vortical structure, which develop separately before their merger.
The region below the leading edge and the suction-side shear layer appear with opposite signs in the third LE mode, suggesting a distinction between the lift-related responses associated with gust impingement and the subsequent development of the suction-side shear layer.
Sign changes along the span appear around the merged LEV in the third BS mode and near the forward suction side in the second LE mode, reflecting the three-dimensionality of the flow seen in Fig.~\ref{fig:cl_three}.

\begin{figure}[!t]
    \centering
    \includegraphics[width=164.9mm]{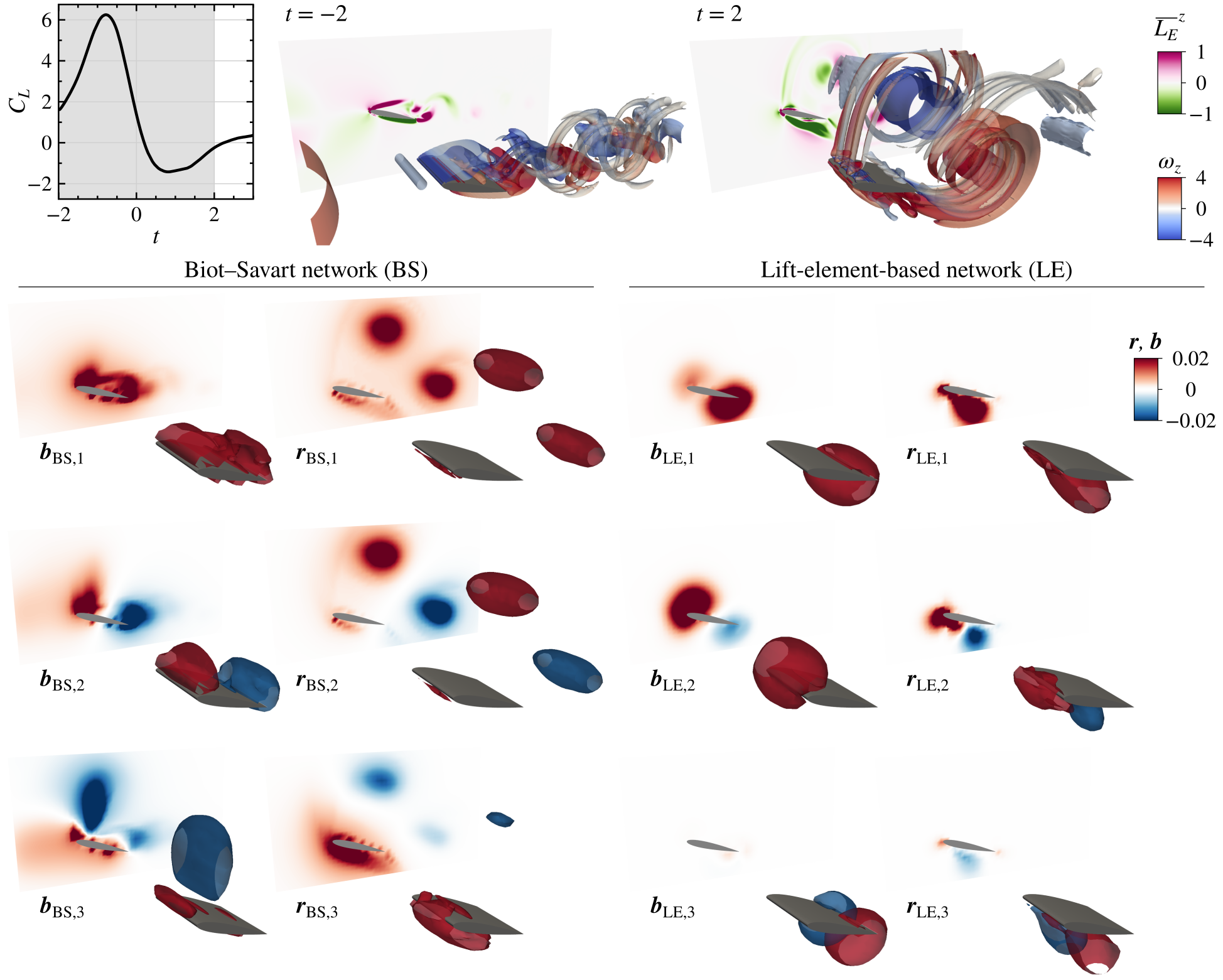}
    \caption{First three modes for $(G, D/c) = (2, 1.5)$.
    Top left: $C_L$, with the analysis window $-2 \le t \le 2$ shaded.
    Top right: flow fields at $t=-2$ and $2$ ($Q$-criterion isosurfaces colored by $\omega_z$; $\overline{L_E}^{\,z}$ on the plane).
    Bottom: forcing modes $\bm{b}_{\mathrm{BS},n}$, $\bm{b}_{\mathrm{LE},n}$ and response modes $\bm{r}_{\mathrm{BS},n}$, $\bm{r}_{\mathrm{LE},n}$ for $n=1$\textmd{--}$3$ (top to bottom), shown by spanwise averages on the plane and isosurfaces at $\pm 0.02$.}
    \label{fig:g2_d1.5_modes}
\end{figure}

\subsubsection{Positive large gust: $(G, D/c) = (2, 1.5)$}

We then examine the effect of gust size, as presented in Fig.~\ref{fig:g2_d1.5_modes}.
We here fix the gust ratio at $G=2$ while varying the gust size $D/c$ from $0.5$ to $1.5$.
For the BS network, the leading forcing mode is located on the forming LEV and TEV around the airfoil, whereas the response mode is concentrated on the two large departing vortices.
Unlike the gust-LEV pair in the case with a small-sized gust, this pair consists of a large LEV and TEV of opposite rotation, whose mutual induction carries them away from the airfoil.

For the LE network, the leading forcing mode lies mostly below the rear pressure side, with a weak part at the leading edge, whereas the response mode extends along the pressure side.
At the end of the analysis window, large negative lift elements extend below the airfoil from the leading edge to the trailing edge, coinciding with the response mode on the pressure-side shear layer thickened to approximately mid-chord, as seen in Fig.~\ref{fig:g2_d1.5_flow}.
The LE pathway thus connects the gust interaction near the rear pressure side to the negative lift contribution associated with the development of the pressure-side shear layer.

The higher modes separate the departing vortices from the regions near the airfoil.
The forming LEV and the TEV side in the forcing, and the upper and downstream departing vortices in the response, appear with opposite signs in the second BS mode, separating the LEV and TEV systems that form on opposite sides of the airfoil.
The leading-edge region dominates the forcing, and the region below the leading edge and the pressure-side shear layer appear with opposite signs in the response of the second LE mode, separating the lift generation near the leading edge from that along the thickened shear layer.
The region above the airfoil, including the departing vortices, is separated from the flow near the airfoil surface in the third BS mode, which is nearly uniform along the span, whereas a sign change along the span appears only along the pressure-side shear layer in the third LE mode, reflecting the three-dimensionality of the flow seen in Fig.~\ref{fig:cl_three}.

\begin{figure}[!t]
    \centering
    \includegraphics[width=164.4mm]{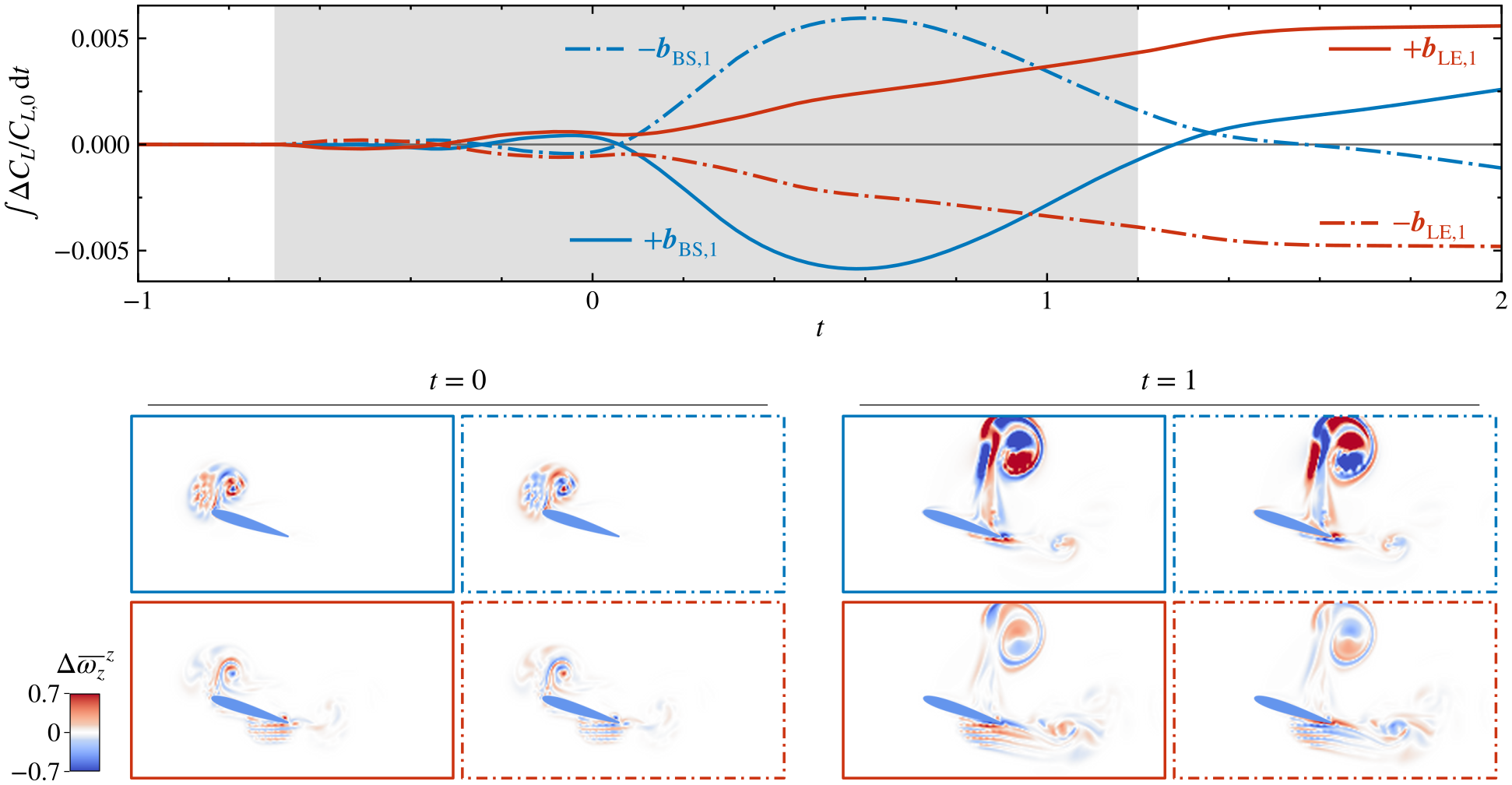}
    \caption{Response to the perturbations along the leading forcing modes, $\pm\bm{b}_{\mathrm{BS},1}$ (blue) and $\pm\bm{b}_{\mathrm{LE},1}$ (red), for $(G, D/c) = (2, 0.5)$; solid and dash-dotted lines and frames denote $+\bm{b}_1$ and $-\bm{b}_1$, respectively. Top: time integral of the lift change, $\int \Delta C_L / C_{L,0}\, \mathrm{d}t$, where $\Delta C_L$ is the difference from the unperturbed lift and $C_{L,0}$ is the lift coefficient at the start of the forcing; the shaded region indicates the analysis window, $-0.7 \le t \le 1.2$. Bottom: difference in the spanwise-averaged spanwise vorticity from the unperturbed flow, $\Delta\overline{\omega_z}^{\,z}$, at $t=0$ (left) and $t=1$ (right) for the BS (upper row) and LE (lower row) networks.}
    \label{fig:omega_l2}
\end{figure}

Across all three cases, the leading BS modes capture vortex development through vortical interactions, including vortex-pair departure and gust-LEV merger, whereas the leading LE modes capture lift-related pathways associated with pressure- or suction-side shear-layer development.
The higher modes provide additional detail on the input-output relationships associated with individual vortical structures and near-airfoil shear layers.

%%%%%%%%%%%%%%%%%%%%%%%%%%%%%%%%%%%%%%%%%%%%%%%%%%%%%%%%%%%%%%

\subsection{Perturbation analysis}\label{sec:perturbation}

To compare the effects of the leading BS and LE forcing modes on the flow and lift, we introduce perturbations based on these modes for $(G, D/c) = (2, 0.5)$.
We apply mode-based vorticity forcing through the right-hand side of the vorticity transport equation~\cite{Yeh2021,Beppu2026},

\begin{equation}
  \frac{\partial \bm{\omega}}{\partial t}
  = \mathcal{N}(\bm{\omega}) + a\bm{b}_f(\bm{x},t),
  \label{eq:forced_vorticity}
\end{equation}
where $\mathcal{N}$ is the unforced three-dimensional vorticity transport operator, $\bm{b}_f$ is the forcing field constructed from the vorticity pulses, and $a$ is the signed forcing amplitude.
In the simulation, this forcing is imposed as a body force in the momentum equations whose curl gives $a\bm{b}_f$.
The leading forcing modes are extracted from the unperturbed flow over short windows of duration $\Delta t_\mathrm{w}=0.1$.
The forcing is applied continuously, with its spatial distribution updated for each short window.
The forcing amplitude is kept small to examine the flow response in the linear regime.

For each short window, $\bm{b}_f$ is constructed by superposing vorticity pulses at the nodes corresponding to the largest 1\% of entries of the leading forcing mode~\cite{Yeh2021}.
The pulses are in the form given in Eq.~(\ref{eq:pulse}) with radius $0.02c$ and are weighted by the corresponding mode entries.
The forcing amplitude is chosen to satisfy
\begin{equation}
  \frac{|a|\,\|\bm{b}_f\|_2}
       {\|\bm{\omega}_{\mathrm{G}}\|_2} = 1,
  \label{eq:amp}
\end{equation}
where $\bm{\omega}_{\mathrm{G}}$ is the vorticity field of the gust at its introduction.
The same forcing norm is prescribed for both networks, and both positive and negative values of $a$ are considered.
Positive and negative forcing are denoted by $+\bm{b}_1$ and $-\bm{b}_1$, respectively.

Figure~\ref{fig:omega_l2} shows the resulting changes in lift and vorticity.
The flow modification is assessed using the vorticity difference $\Delta\bm{\omega}$ from the unperturbed flow.
The lift response is assessed using the time integral $\int_{t_0}^{t} \Delta C_L / C_{L,0}\,\mathrm{d}t'$, where $\Delta C_L$ is the change from the unperturbed lift coefficient and $C_{L,0}$ is the lift coefficient at $t_0 = -0.7$, when the forcing is started.

The BS forcing primarily modifies the gust-LEV pair, whereas the LE forcing mainly modifies the pressure-side shear layer.
At $t = 0$, the BS-induced vorticity change is concentrated around the gust and the LEV near the leading edge, while the LE-induced change also extends along the pressure-side shear layer near the trailing edge.
At $t = 1$, the BS-induced change follows the departing pair, whereas the LE-induced change remains concentrated in the pressure-side shear layer and trailing-edge wake, with a weaker change in the departing pair.

The lift responses to positive and negative forcing are nearly antisymmetric for both networks, consistent with an approximately linear response at this amplitude.
Compared with the lift history in Fig.~\ref{fig:g2_d0.5_modes}, the $+\bm{b}_{\mathrm{BS},1}$ forcing amplifies the gust-induced lift excursion by increasing the lift maximum and decreasing the lift minimum, whereas the $-\bm{b}_{\mathrm{BS},1}$ forcing attenuates it.
The slope of the time integral gives the instantaneous lift change, and its extrema at $t \approx 0$ and $0.6$, where the BS-induced lift change reverses sign, appear shortly after the lift begins to decrease at $t \approx -0.2$ and to recover at $t \approx 0.4$, respectively.
The BS-induced lift change thus follows the lift variation driven by the gust, indicating that the BS forcing modifies the gust-airfoil interaction itself.
In contrast, the LE forcing increases or decreases both the lift maximum and minimum, depending on the forcing sign, and produces a sustained lift increase or decrease.
This difference in lift response corresponds to the vortex pair and near-airfoil shear layer emphasized by the BS and LE response modes, respectively.

%%%%%%%%%%%%%%%%%%%%%%%%%%%%%%%%%%%%%%%%%%%%%%%%%%%%%%%%%%%%%%
%%%%%%%%%%%%%%%%%%%%%%%%%%%%%%%%%%%%%%%%%%%%%%%%%%%%%%%%%%%%%%
\section{Conclusion}\label{sec:conclusion}

We introduced a lift-element-based network and compared its broadcast modes with those of a Biot--Savart network for three vortex gust-airfoil interactions at $Re = 1000$.
The leading BS response modes emphasized the principal vortices formed during the gust encounters, whereas the LE response modes emphasized the near-airfoil shear layers on the pressure or suction side.
Higher modes distinguished vortex-formation regions and detached versus near-airfoil structures, and revealed spanwise variations in the forcing and response patterns.
Perturbation analysis supported this modal interpretation: forcing based on the leading BS modes amplified or attenuated the gust-induced lift variation, whereas forcing based on the LE modes produced a sustained lift increase or decrease.

By incorporating force elements into the network weights, the proposed framework extracts modes that directly characterize causal input-output relationships between local flow structures and transient lift.
These modes provide a physical basis for understanding how aerodynamic loads arise and evolve in strongly unsteady flows.
The present framework can be extended to other aerodynamic loads, such as drag and moment, by changing the auxiliary potential.
For the present flows, the drag response is similar to the lift response~\cite{fukami2025extreme}, whereas a drag-element-based network may be useful for aerodynamic flows in which the lift and drag mechanisms are coupled but differ from each other.
The forcing modes can guide the placement and timing of actuation for mitigating transient aerodynamic loads in extreme gust encounters.

%%%%%%%%%%%%%%%%%%%%%%%%%%%%%%%%%%%%%%%%%%%%%%%%%%%%%%%%%%%%%%
%%%%%%%%%%%%%%%%%%%%%%%%%%%%%%%%%%%%%%%%%%%%%%%%%%%%%%%%%%%%%%
%%%%%%%%%%%%%%%%%%%%%%%%%%%%%%%%%%%%%%%%%%%%%%%%%%%%%%%%%%%%%%
\section*{Acknowledgments}

Kai Fukami acknowledges support from the JSPS KAKENHI grant nos. JP25K23418 and JP26K01129, the JST PRESTO grant no. JPMJPR25KA, and the MEXT Coordination Funds for Promoting Aerospace Utilization grant no. JPJ000959.
Part of this work was carried out using computational resources of the supercomputer Grand Chariot 2 provided by Hokkaido University through the HPCI System Research Projects (Project ID: hp260005).

% \section*{Acknowledgments}
% {
% {
% Kai Fukami acknowledges support from the JSPS KAKENHI Grant No. JP25K23418 and No. JP26K01129, the JST PRESTO Grant No. JPMJPR25KA, and the MEXT Coordination Funds for Promoting Aerospace Utilization Grant No. JPJ000959.
% Yuzuru Kato acknowledges support from the JST PRESTO Grant No. JPMJPR24K3.
% }}

% \vspace{-5mm}
\section*{Declaration of interests}
{The authors report no conflict of interest.}

%%%%%%%%%%%%%%%%%%%%%%%%%%%%%%%%
%%%%%%%%%%%%%%%%%%%%%%%%%%%%%%%%%
%%%%%%%%%%%%%%%%%%%%%%%%%%%%%%%%%

\bibliographystyle{unsrt}  
\bibliography{refs}
%%% Remove comment to use the external .bib file (using bibtex).

\end{document}